\documentclass[10pt]{iopart}

\usepackage{iopams}  
\usepackage{color, soul}
\usepackage{graphicx}
\usepackage{float}
\usepackage{times,mathptmx}
\usepackage{array}
\expandafter\let\csname equation*\endcsname\relax

\expandafter\let\csname endequation*\endcsname\relax

\usepackage{breqn}

\begin{document}

\title {{Spin pumping and inverse spin Hall effect in CoFeB/C$_{60}$ bilayers}}

\author{Purbasha Sharangi$^1$, Braj Bhusan Singh$^1$, Sagarika Nayak $^1$, Subhankar Bedanta$^{1,2}$}

\address{$^1$ Laboratory for Nanomagnetism and Magnetic Materials (LNMM), School of Physical Sciences, National Institute of Science Education and Research (NISER), HBNI, P.O.- Bhimpur Padanpur, Via Jatni, 752050, India}
\address{$^2$ Center for Interdisciplinary Sciences (CIS), National Institute of Science Education and Research (NISER), HBNI, Jatni, 752050 India}

\ead{sbedanta@niser.ac.in}
\vspace{10pt}
\begin{indented}
\item[]July 2021
\end{indented}

\begin{abstract}
Pure spin current based research is mostly focused on ferromagnet (FM)/heavy metal (HM) systems. Because of the high spin orbit coupling (SOC) these HMs exhibit short spin diffusion length and therefore possess challenges for device application. Low SOC (elements of light weight) and large spin diffusion length make the organic semiconductors (OSCs) suitable for future spintronic applications. It has been predicted that curvature may enhance SOC of C$_{60}$ due to hybridization of the $\pi$ - $\sigma$ electrons of carbon atoms. The presence of reasonable SOC can be contemplated via spin pumping experiments in FM/C$_{60}$ heterostructure. In this context, we have investigated spin pumping and inverse spin hall effect (ISHE) in CoFeB/C$_{60}$ bilayer system using coplanar wave guide based ferromagnetic resonance (CPW-FMR) set-up. We have performed angle dependent ISHE measurement to disentangle the spin rectification effects for example anisotropic magnetoresistance, anomalous Hall effect etc. Further, effective spin mixing conductance (g$_{eff}^{\uparrow\downarrow}$) and spin Hall angle ($\theta_{SH}$) for C$_{60}$ have been reported here. The evaluated value for $\theta_{SH}$ is 0.059. 
\end{abstract}

%
\ioptwocol

\section{Introduction:}
	Pure spin current plays an important role in fabrication of power efficient spintronics devices \cite{tserkovnyak2005nonlocal}. Spin pumping is very efficient way of creation of pure spin current in ferromagnetic (FM)/nonmagnetic (NM) system by microwave excitation which can be converted into a voltage by inverse spin Hall effect (ISHE) \cite{hirsch1999spin,tserkovnyak2002spin,valenzuela2006direct}. Various materials possessing high spin orbit coupling (SOC) like Pt, Ta, W, Bi$_{2}$Se$_{3}$ etc., have been utilized for efficient spin to charge current conversion \cite{niimi2015reciprocal,singh2019inverse,singh2017study,singh2021high}. Due to high SOC, the spin life time and spin diffusion length are less for heavy metals (HM) \cite{wolf2001spintronics,vzutic2004spintronics}. Therefore, there is a search for new materials which exhibit large spin diffusion length ($\lambda_{sd}$) for such spintronic applications. In this context organic semiconductors (OSCs) have shown the potential due to low SOC, less hyperfine interaction (HFI) and long $\lambda_{sd}$ \cite{sun2014first,dediu2009spin,atodiresei2010design,barraud2010unravelling}. OSCs are extensively studied in spin valve (SV), organic light emitting diode (OLED) and spin optical devices \cite{zhang2013observation,liu2018studies,watanabe2014polaron,liang2016curvature,gobbi2011room}. Among all OSCs, C$_{60}$ is emerging as a potential candidate due to the absence of hydrogen atom and negligible HFI. FM/C$_{60}$ interface is very crucial in spin polarized charge transfer. C$_{60}$ can exhibit ferromagnetism at the interface (known as spinterface) when it is deposited on any ferromagnet \cite{sanvito2010molecular,djeghloul2013direct}. Presence of C$_{60}$ can modify the magnetic property of the ferromagnetic materials as well as nonmagnetic metal \cite{moorsom2014spin,mallik2018effect,mallik2019tuning,mallik2019enhanced,al2015beating,sharangi2021magnetism}. It is noted that the low value of SOC in OSC limits them for efficient spin to charge current conversion. However, it has been reported that when C$_{60}$ is deposited on a substrate it may exhibit enhancement of SOC \cite{liang2016curvature,alotibi2021enhanced}. There are very few reports on generation of pure spin current in FM/OSC system \cite{liu2018studies,sun2016inverse,das2018enhanced,liu2019studies,kalappattil2020giant}. Inserting a small C$_{60}$ layer can reduce the conductivity mismatch between YIG and Pt \cite{das2018enhanced,kalappattil2020giant}. The presence of C$_{60}$ also decrease perpendicular magnetic anisotropy (PMA) of YIG and enhance spin mixing conductance across YIG/C$_{60}$/Pt interfaces \cite{kalappattil2020giant}. D. Sun et al., have studied pulsed ISHE response in a variety of OSCs and quantitively determined the spin diffusion length and spin Hall angle \cite{sun2016inverse}. Using the pulsed FMR technique one can generate two to three order larger magnitude of ISHE signal compared to normal microwave excitation in a coplanar waveguide (CPW).  They have shown that among all OSCs, C$_{60}$ possess larger spin current signal and the evaluated spin Hall angle ($\theta_{SH}$) for C$_{60}$ is 0.014. Further, Liu et al., have observed spin pumping in NiFe/C$_{60}$/Pt system \cite{liu2018studies}. To detect the spin current signal, they have used Pt layer (which has high SOC). However, in this paper we have investigated the spin pumping and ISHE in a bilayer CoFeB/C$_{60}$ system using simple microwave excitation in a CPW. 
	CoFeB is a low Gilbert damping material which is an important parameter from application point of view. Also, it is amorphous in nature and preparation of such amorphous thin film is relatively easy. It should be noted that here we have not used any HM layer like Pt. Angle dependent ISHE measurements have been performed to know the individual contribution of spin pumping and spin rectification effects. We have also calculated the spin mixing conductance and spin Hall angle for C$_{60}$.

\section{Experimental details:}

We have prepared CoFeB (5 nm)/C$_{60}$ bilayer samples with varying the thickness of C$_{60}$ ($t_{C_{60}}$). The samples are named as S1 to S5 having $t_{C_{60}}$  = 0, 1.1, 2.1, 5.2 and 14.8 nm, respectively. The detailed sample structure is shown in fig. 1(a). The samples have been prepared on Si(100) /native oxide substrate using DC sputtering (CoFeB), thermal evaporation (C$_{60}$) and e-beam evaporation (MgO) techniques in a multi-deposition high vacuum chamber manufactured by Mantis Deposition Ltd., UK. The base pressure of the vacuum chamber was ~5$\times$$10^{-8}$ mbar. All the layers were deposited without breaking the vacuum to avoid oxidation and surface contamination. The deposition pressure was 5$\times$$10^{-3}$ mbar for CoFeB and 1$\times$$10^{-7}$ mbar for C$_{60}$ and MgO evaporation. The deposition rates for CoFeB and C$_{60}$ layers were 0.1 and $\sim$ 0.15  Å/sec, respectively. 2 nm of MgO layer on top of C$_{60}$ layer with deposition rate $\sim$ 0.05  Å/sec has been deposited. 
Cross-sectional transmission electron microscopy (TEM) has been performed on sample S4 using a high-resolution TEM (JEOL F200, operating at 200 kV and equipped with a GATAN oneview CMOS camera).

We have performed ferromagnetic resonance (FMR) measurement for all the samples in a frequency range 6 - 17 GHz. During the measurement the sample was kept on the wide coplanar waveguide (CPW) in the flip-chip manner. Angle dependent ISHE measurements have been performed at a fixed frequency of 7 GHz and 11 mW power by connecting a nano-voltmeter (Keithley 2182A) at the opposite edges of the sample using silver paste contacts. The measurement geometry of the FMR and ISHE are shown in Fig. 1 and the detailed methodology have been discussed in our earlier reports \cite{singh2019inverse,singh2017study}.

\section{Results and discussion:}
 
 \begin{figure}
 	\centering
 	\includegraphics[width=1.0\linewidth]{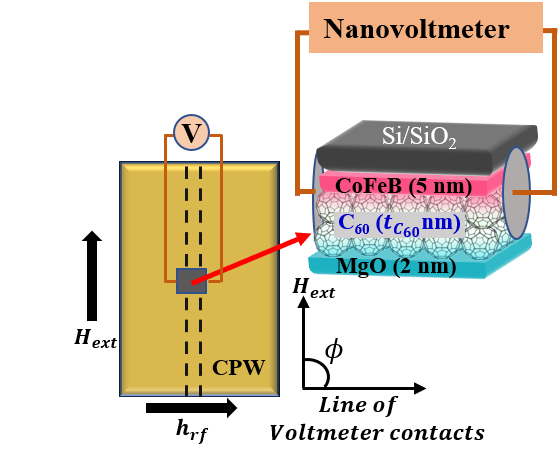}
 	\caption{Schematic of the sample structure, FMR and ISHE voltage measurement geometry. The thicknesses shown in this schematic is not to scale to the exact thicknesses of the samples. }
 	\label{fig:fig1}
 \end{figure}
  
 For structural characterization we have performed cross-sectional TEM. The high resolution TEM image shows the amorphous growth of CoFeB and C$_{60}$ layers\cite{doi:10.1021/acs.jpcc.1c08656}. The sharp interface indicates that there is no interdiffusion between CoFeB and C$_{60}$ layers\cite{doi:10.1021/acs.jpcc.1c08656}.

In order to evaluate the magnetic damping constant ($\alpha$), frequency dependent FMR measurement has been performed in the frequency range 6 – 17 GHz (FMR spectra are shown in supplementary figure S1). The evaluated values of $\alpha$ for samples S1 to S5 are 0.0094 $\pm$ 0.0001, 0.0106 $\pm$ 0.0002, 0.0110 $\pm$ 0.0002, 0.0124 $\pm$ 0.0003 and 0.0169± 0.0006, respectively \cite{doi:10.1021/acs.jpcc.1c08656}. The increase in damping may be due to the interface roughness or some other effect at interface. This enhancement in $\alpha$ in the bilayer (CoFeB/C$_{60}$) samples as compared to the single layer (without C$_{60}$) indicates that there is a possibility of spin pumping from CoFeB to C$_{60}$. 

In order to evaluate the presence of spin pumping in CoFeB/C$_{60}$ interface we have performed the in-plane angle ($\phi$) dependent ISHE measurement in all the samples. Figure 2 shows measured ISHE voltage (V$_{meas}$) and FMR signal vs applied magnetic field (H) plots for the sample S5 at 0$^{\circ}$. Here $\phi$ denotes the angle between the line of contact pad and applied DC magnetic field ($H$). Since, $V_{meas}$ $\propto$ ($H$ $\times$ $\sigma$), where $\sigma$ is the direction of spin polarization, upon changing the magnetic field ($H$) direction, $V_{meas}$ also changes. ISHE signal changes its polarity from 0$^{\circ}$ to 180$^{\circ}$ as the magnetic field direction is reversed, which indicates that the primary signal is coming from the spin pumping at CoFeB/C$_{60}$ interface (data not shown). We have not observed any ISHE signal in CoFeB single layer (supplementary figure S2). Therefore, the observed ISHE signal is due to spin pumping at CoFeB-C$_{60}$ interface. Due to the hybridization of $\pi$ - $\sigma$  electrons of carbon atoms, there is a probability of curvature enhanced SOC in C$_{60}$, which leads to the spin pumping at CoFeB/C$_{60}$ interface. Spin accumulation is negligible when $\phi$ = 90$^{\circ}$, which show negligible ISHE signal for all the samples having C$_{60}$(data not shown). 

To separate the symmetric (V$_{sym}$) and anti-symmetric (V$_{asym}$) components from the ISHE signal, we have fitted the experimental data using the following Lorentzian equation \cite{iguchi2017measurement}:

\begin{eqnarray}
	\begin{aligned}
		V_{meas} = V_{sym} \frac{(\Delta H)^2}{(H-H_r)^2+(\Delta H)^2}+ \\
		V_{asym} \frac{2 \Delta H (H - H_r)}{(H-H_r)^2+(\Delta H)^2}
	\end{aligned}
\end{eqnarray}

\begin{figure}
	\centering
	\includegraphics[width=1.0\linewidth]{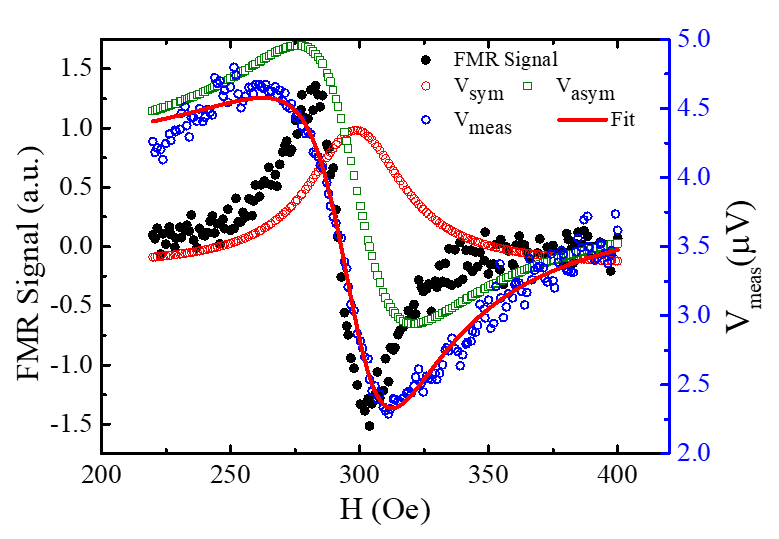}
	\caption{$V_{meas}$ (blue open circle) and FMR signal (black solid circle) vs applied magnetic field ($H$) for the sample S5 at 0$^{\circ}$. Solid red lines are the best fit to the eq. (1), while green open square and red open circle represent the $V_{asym}$ and $V_{sym}$ contributions of the ISHE voltage, respectively.}
	\label{fig:figure-2}
\end{figure}

\begin{figure*}
	\centering
	\includegraphics[width=0.81\linewidth]{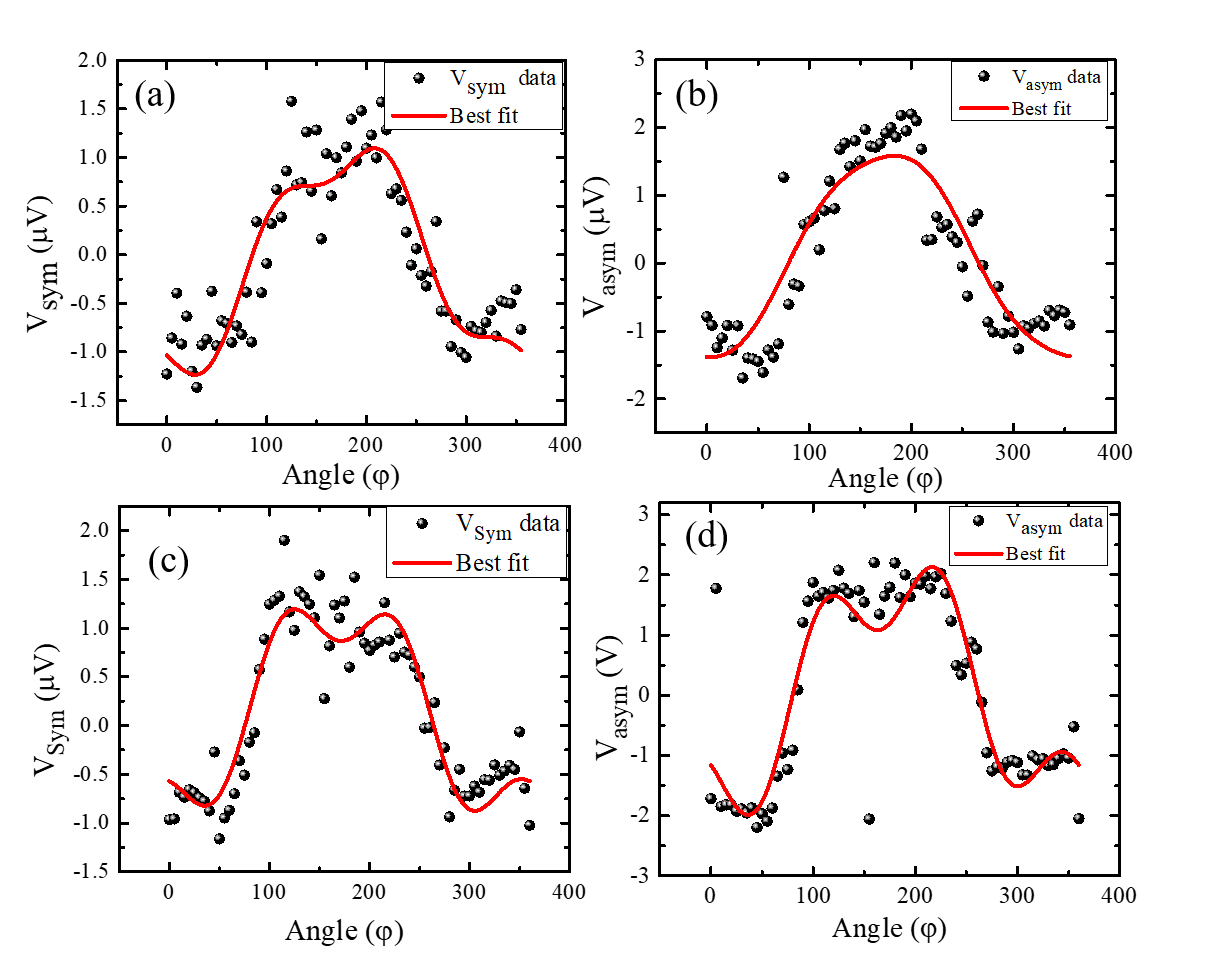}
	\caption{Angle dependency of evaluated ISHE voltage components Vsym and Vasym for samples S3 (a and b) and S5 (c and d), respectively (open circles). The solid lines are the best fits using eqs. 2 and 3.}
	\label{fig:figure-3}
\end{figure*}

The solid red lines are the best fit to the eq. (1), while green open square and red open circle represent the V$_{asym}$ and V$_{sym}$ contributions, respectively, for sample S5. The  V$_{sym}$ component has the major contribution in spin pumping, while minor contributions in anomalous Hall effect (AHE), and anisotropic magnetoresistance (AMR). In other side  V$_{asym}$ component has the major contribution in AHE and AMR. To quantify the spin pumping voltage (V$_{SP}$) and the other rectification effects i.e., V$_{AHE}$, V$_{AMR}$ etc., we have performed the in-plane angle dependent ISHE measurement with 5$^{\circ}$ interval. Figure 3 shows the V$_{sym}$ and V$_{asym}$ vs $\phi$ plots for sample S3 and S5. The red solid lines are the best fits which are fitted by eq. 2 (for symmetric part) and 3 (antisymmetric part) \cite{conca2017lack}. The angle $\phi$ is defined as the relative phase between the microwave electric and magnetic fields in the medium and in case of our measurement  $\Phi$=90$^{\circ}$. 

\begin{eqnarray}
	\begin{aligned}
		V_{sym}= V_{sp}cos^3(\phi)+ V_{AHE}cos(\phi)cos(\Phi)+ \\ V_{sym}^{AMR \perp} cos(2\phi)cos(\phi)+ V_{sym}^{AMR ||}sin(2\phi)cos(\phi)
	\end{aligned}
\end{eqnarray}

\begin{eqnarray}
	\begin{aligned}
		V_{asym}= V_{AHE} cos(\phi)sin(\Phi) + V_{asym}^{AMR \perp} cos(2\phi)sin(\phi)+ 
		\\ V_{asym}^{AMR ||}sin(2\phi)cos(\phi)
	\end{aligned}
\end{eqnarray}

The obtained values of $V_{SP}$, $V_{AHE}^{aSym}$, $V_{AMR}^{\perp}$, $V_{AMR}^{\parallel}$ from the fitted data are listed in table 1.

\begin{table*}
	\caption{Fitted parameters from angle dependent ISHE measurements for sample S2 to S5 }

	\centering
	\begin{tabular}{c c c c c c c} 
		
		Sample&  $V_{SP}$ ($\mu$V)& $V_{Sym}^{AMR\perp}$ ($\mu$V)& $V_{Sym}^{AMR\parallel}$ ($\mu$V)&$V_{AHE}^{aSym}$ ($\mu$V)&$V_{aSym}^{AMR\perp}$ ($\mu$V)&$V_{aSym}^{AMR\parallel}$ ($\mu$V) \\ \hline
		S2     & -2.34 $\pm$ 0.11              & 1.54 $\pm$ 0.08              & 0.15 $\pm$ 0.06                       & -0.99 $\pm$ 0.07       & 0.24 $\pm$ 0.11 & 0.25 $\pm$ 0.08                 \\ \hline
		S3     & -2.66 $\pm$ 0.20              & 1.76 $\pm$ 0.21              & -0.25 $\pm$ 0.11                       & -1.60 $\pm$ 0.15       & 0.19 $\pm$ 0.13 & -0.19 $\pm$ 0.09                \\ \hline
		S4     & -2.68 $\pm$ 0.13              & 1.36 $\pm$ 0.14              & 0.10 $\pm$ 0.08                       & -2.08 $\pm$ 0.21       & 0.37 $\pm$ 0.25 & 0.36 $\pm$ 0.25              \\ \hline
		S5     & -2.85 $\pm$ 0.15              & 2.13 $\pm$ 0.16              & 0.07 $\pm$ 0.05                       & -2.55 $\pm$ 0.16       & 1.48 $\pm$ 0.23 & -0.35 $\pm$ 0.16           \\ \hline
	\end{tabular}
	
\end{table*}

It has been observed that spin pumping voltage increases with increasing the thickness of  C$_{60}$.
The effective spin mixing conductance (g$_{eff}^{\uparrow\downarrow}$) plays an important role in spin transport phenomenon for FM/NM interface. One can calculate the parameter using the below eq. \cite{singh2020high}:
\begin{eqnarray}
	\begin{aligned}
		g_{eff}^{\uparrow\downarrow}= \frac{\Delta\alpha4\pi M_{S}t_{FM}}{g\mu_B}
	\end{aligned}
\end{eqnarray}

Where $\Delta\alpha$, $t_{FM}$, $\mu_B$, g are the change in $\alpha$ due to spin pumping, the thickness of CoFeB layer, Bohr magneton, and Lande’s g- factor ($\sim$2.5), respectively. Using SQUID magnetometry, we have measured the value of $M_{S}$, which is 1015, 990, 970, 960 and 945 emu/cc for samples S1 to S5, respectively. The calculated g$_{eff}^{\uparrow\downarrow}$ values for samples S2 to S5 are  3.24$\times$$10^{18}$, 4.22$\times$$10^{18}$,7.55$\times$$10^{18}$and 1.72$\times$$10^{19}$ $m^{-2}$. It is observed that g$_{eff}^{\uparrow\downarrow}$ increases with increasing the $t_{C_{60}}$. The possible reason behind it may be the reduction in interface roughness by the well growth of C$_{60}$ and also the formation of spinterface with enhanced hybridization in higher $t_{C_{60}}$.

Further, we have calculated the spin Hall angle (SHA) and spin diffusion length ($\lambda_{C_{60}}$) for  C$_{60}$ using a spin diffusion model given by the following equation \cite{singh2020high}.  

\begin{eqnarray}
	\begin{aligned}
		\frac {V_{sp}} {R}= (w)\times \theta_{SH} \lambda_{C_{60}}tanh(\frac{t_{C_{60}}}{2 \lambda_{C_{60}}}) |\Vec{J}_s|
	\end{aligned}
\end{eqnarray}
where,

\begin{eqnarray}
	\begin{aligned}
		|\Vec{J}_s| \approx (\frac{g_{r}^{\uparrow \downarrow}\hbar}{8\pi})(\frac{\mu_0 h_{rf}\gamma}{\alpha})^2\times\\
		[\frac{\mu_0 M_s\gamma+\sqrt{(\mu_0 M_s\gamma)^2+16(\pi f)^2}}{(\mu_0 M_s\gamma)^2+16(\pi f)^2}](\frac{2e}{\hbar})
	\end{aligned}
\end{eqnarray}

\begin{figure}[H]
	\centering
	\includegraphics[width=1.1\linewidth]{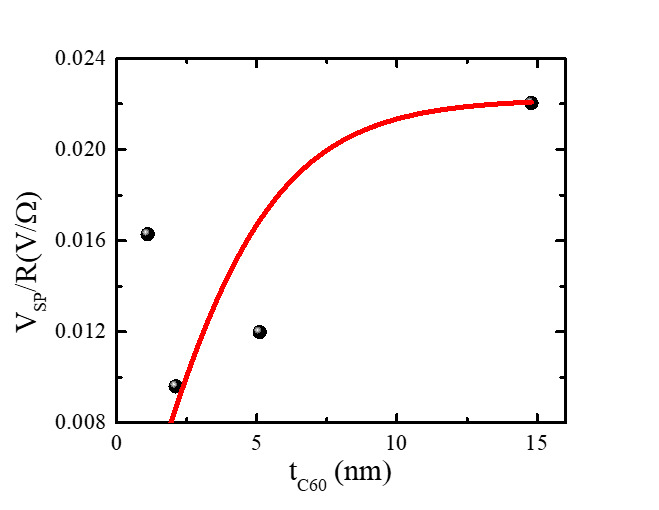}
	\caption{Thickness ($t_{C_{60}}$) dependence of normalized spin pumping voltage V$_{SP}$/R. Black solid circles are the experimental data and the red solid line is the best fit using eq. 5.}
	\label{fig:figure-4}
\end{figure}

The resistivity of the samples was measured using four probe technique. R is the resistance of the samples. The values of the CPW transmission line width (w) and rf field ($\mu_{0} h_{rf}$) are 200 $\mu$m and 0.05 mT, respectively. 
Figure 4 shows the change in V$_{SP}$/R with $t_{C_{60}}$. The black solid circles are the experimental data and the red solid line is the best fit using eq. (5). From the best fit $\lambda_{sd}$ of  C$_{60}$ for CoFeB/ C$_{60}$ system and the product of $\theta_{SH}$\textit{$g_{\it{eff}}^{\uparrow \downarrow}$} are found to be 2.56 nm and 1.01$\times$10$^{18 }$ m$^{-2}$, respectively. The SHA of the sample is found to be 0.059 which is higher than the reported value by Sun et al., \cite{sun2016inverse}.

\section{Conclusions:} 

In summary, we have performed a detail investigation of spin pumping and ISHE in CoFeB/C$_{60}$ system with different C$_{60}$ thicknesses. Here, CPW based FMR has been used to detect the ISHE signal. The origin of this spin pumping at CoFeB/C$_{60}$ interface is probably due to curvature enhanced SOC in C$_{60}$. Damping, spin pumping voltage and effective spin mixing conductance show monotonic increment with C$_{60}$ thickness. The obtained spin Hall angle for CoFeB/C$_{60}$ bilayer system is 0.059 which is higher than the reported value. Future works are needed to understand the underneath mechanism of this curvature enhanced SOC and spin pumping at FM/OSC interface.

\section{Acknowledgments:}
The authors want to thank Dr. Tapas Gosh and Mr. Shaktiranjan Mohanty for helping in TEM imaging. The authors also acknowledge Department of Atomic Energy, and Department of Science and Technology - Science and Engineering Research Board, Govt. of India (DST/EMR/2016/007725) for the financial support.

\section{References:}

\bibliography{reference}

\end{document}


\maketitle
\onecolumngrid

\LARGE{\textbf{Supplementary Information \\Spin pumping and inverse spin Hall effect in CoFeB/C$_{60}$ bilayers}} \\

\large{Purbasha Sharangi\textit{$^{1}$}, Braj Bhusan Singh\textit{$^{1}$}, Sagarika Nayak\textit{$^{1}$}, and Subhankar Bedanta$^{\ast}$\textit{$^{1,2}$}}\\

\large{\textit{$^{1}$~Laboratory for Nanomagnetism and Magnetic Materials (LNMM), School of Physical Sciences, National Institute of Science Education and Research (NISER), HBNI, P.O.- Bhimpur Padanpur, Via-Jatni, 752050, India. E-mail: sbedanta@niser.ac.in}}\\
\large{\textit{$^{2}$~Center for Interdisciplinary Sciences (CIS), National Institute of Science Education and Research (NISER), HBNI, Jatni, 752050 India}}\\

FMR spectra measured at 7 GHz frequency for all the samples are shown in figure S1. The measured spectra are fitted with the following Lorentzian expression:
\begin{eqnarray}
	\begin{aligned}
		FMR \ signal = K_{1} \frac{4\Delta H(H-H_r)}{4(H-H_r)^2+(\Delta H)^2}+ 
		K_{2} \frac{(\Delta H)^2-4(H - H_r)^2}{4(H-H_r)^2+(\Delta H)^2}
	\end{aligned}
\end{eqnarray}

\begin{figure}[H]
	\centering
	\includegraphics[width=0.7\linewidth]{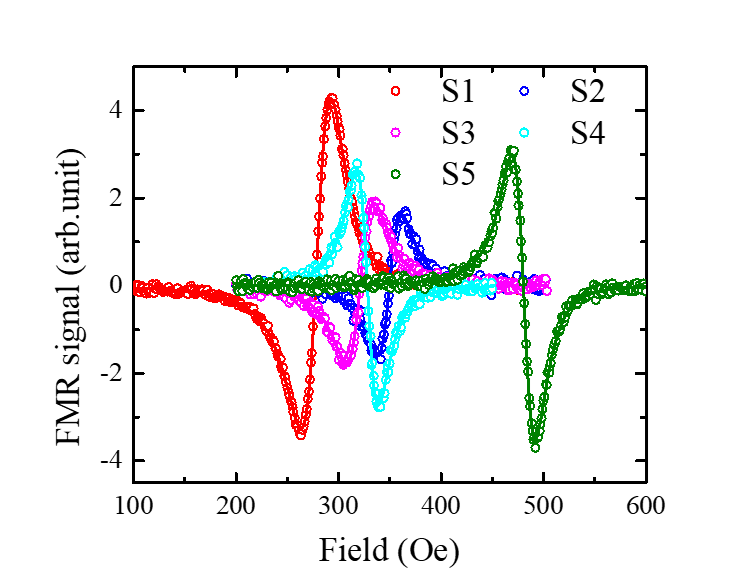}
	\caption{FMR spectra measured by sweeping the dc magnetic field ($H$) at 7 GHz frequency for all the samples. } \label{fig:fig-S1}
\end{figure}
 
 Fig. S2 shows FMR signal and $V_{meas}$ vs applied magnetic field ($H$) for sample S1 (5 nm CoFeB layer). Single layer CoFeB layer exhibit no ISHE signal, which confirms that the measured ISHE signal for CoFeB/C$_{60}$ samples are due to the spin pumping at FM/OSC interface.
 
\begin{figure*}
	\centering
	\includegraphics[width=0.8\linewidth]{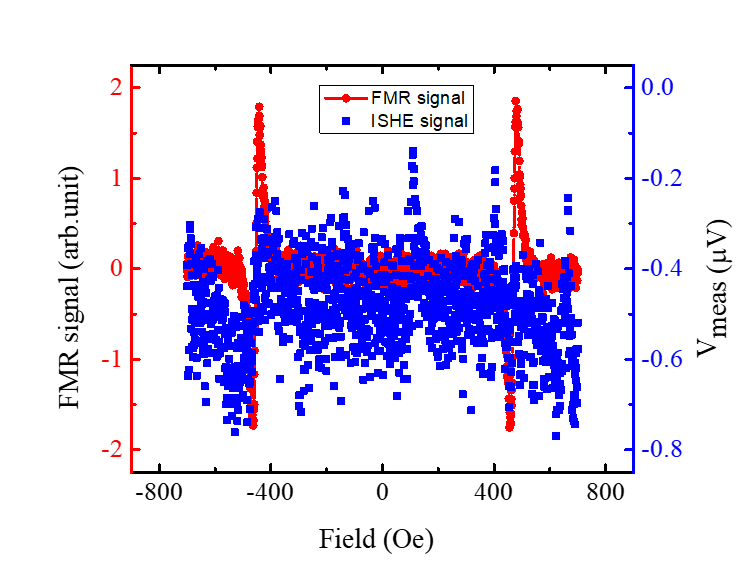}
	\caption{FMR signal and $V_{meas}$ measured by sweeping the dc magnetic field for CoFeB (5 nm) thin film.}	\label{figS2}
\end{figure*}
